# Observation of non-Hermitian antichiral edge currents

Rui Ye[1], Yanyan He[1*], Guangzhen Li[1], Luojia Wang[1], Xiaoxiong Wu[1], Xin Qiao[1], Yuanlin Zheng[1], Liang Jin[2], Da-Wei Wang[3], Luqi Yuan[1*], and Xianfeng Chen[1,4,5*]

[1]*State Key Laboratory of Advanced Optical Communication Systems and Networks, School of Physics and Astronomy, Shanghai Jiao Tong University, Shanghai, 200240, China*
[2]*School of Physics, Nankai University, Tianjin 300071, China*
[3]*Interdisciplinary Center for Quantum Information and Zhejiang Province Key Laboratory of Quantum Technology and Device, Department of Physics, Zhejiang University, Hangzhou 310027, China*
[4]*Shanghai Research Center for Quantum Sciences, Shanghai, 201315, China*
[5]*Collaborative Innovation Center of Light Manipulations and Applications, Shandong Normal University, Jinan, 250358, China*

E-mail: yanyhe@sjtu.edu.cn; yuanluqi@sjtu.edu.cn; xfchen@sjtu.edu.cn
These authors contributed equally: Rui Ye, Yanyan He, and Guangzhen Li

**Abstract**
Non-Hermitian topological photonics is of great interest in bridging topological matter with gain/dissipation engineering in optics. A key problem in this direction is the interplay between the effective gauge potential and the non-Hermiticity. Here we tackle this problem in a synthetic non-Hermitian Hall ladder and experimentally observe antichiral edge currents (ACECs) of photons, by tuning the locally uniform effective magnetic flux and the on-site gain/loss. Such ACECs provide a topological method to probe the signatures of the non-Hermitian skin effect (NHSE) from steady-state bulk dynamics. The universality of this method is verified by its generalization to three dimensions. This study paves a way to investigate exotic non-Hermitian topological physics and has potential applications in topological photonics engineering.

**Introduction**
Chiral edge currents (CECs) at the boundaries of two-dimensional (2D) topological materials are featured by robust unidirectionality against defects [1-3] and have been demonstrated in Hermitian lattices with effective magnetic fields [4-7]. Intriguingly, the broken chirality in a topological system leads to antichiral edge currents (ACECs) [8-14], which results in edge currents propagating in the same direction. However, in Hermitian systems such chiral symmetry breaking in edge currents requires locally non-uniform magnetic flux in a modified Haldane model [15-17], which is challenging for optical frequency. On the other hand, the gain and loss engineering has recently been developed to demonstrate various intriguing properties of non-Hermitian Hamiltonians [18-27], such as non-reciprocal light propagation [19] and unidirectional invisibility [20], proving to be a powerful tool in controlling the direction of energy flow in physical systems. Hence the non-Hermiticity provides a promising way to seek the achievement of ACECs.

In this work, we experimentally demonstrate non-Hermitian ACECs resulting from the interplay between on-site gain/loss effects and effective magnetic flux for light in optical frequency, fundamentally different from microwave Hermitian ACECs [15-17]. The observed non-Hermitian

ACECs exhibit intrinsic signatures of the non-Hermitian skin effect (NHSE) [28-45], which is observed from steady-state bulk dynamics without obtaining the Lyapunov exponent [46,47]. Such intrinsic correspondence between the ACECs and the NHSE is further verified by measuring the nontrivial topological windings of band structures in the complex energy plane [48-52]. The universality of this strategy is verified by extending the model to a three-dimensional (3D) synthetic lattice and studying the non-Hermitian ACECs with a spectral winding number $-2$. Our experiment is based on the concept of synthetic frequency dimension [53-62], which has been successful in studying various physics including CECs in the quantum Hall ladder [56] and measuring topological windings [57]. We use two coupled ring resonators under independent dynamic modulations to create a Hall ladder with locally uniform effective magnetic flux and apply a third modulation to add the on-site gain/loss. Our results hence realize optical ACECs from non-Hermitian topology with a protocol possibly generalized to all the electromagnetic wavelengths, which may also have potential applications with unidirectional frequency conversions near the telecom frequency.

## Results

**Correspondence between ACECs and the NHSE.** To highlight the underlying physics, we discuss a non-Hermitian two-leg Hall ladder model with an effective magnetic flux $\phi$ in each unit plaquette as well as on-site gain $(i\gamma)$ and loss $(-i\gamma)$ on each leg [see Fig. 1(a)]. In a Hermitian Hall ladder ($\gamma = 0$), there exhibits CECs, i.e., the edge states on either leg ($a$ or $b$) propagate in opposite directions [see the dashed lines in Fig. 1(a)]. However, once gain/loss is added ($\gamma \neq 0$), the currents on two legs propagate in the same direction, which are ACECs [see the solid lines in Fig. 1(a)] and link the NHSE. We first analyze the NHSE in this non-Hermitian quantum Hall ladder model using the standard procedure on solving the relevant Hamiltonian of the model (see Methods), and then give the quantitative correspondence between the ACECs and NHSE.

In Fig. 1(c), we show the energy spectra with the PBC and OBC in the complex energy plane for different effective magnetic flux $\phi$, where the values of spectral winding number $w$ (see Methods) are given inside each loop. Here the OBC is provided by using a finite lattice with 300 sites. We see the energy spectra under the OBC and PBC are different for cases of $\phi = 0.25\pi, 0.5\pi$, and exhibit $w \neq 0$, indicating the existence of the NHSE [48-52]. Specifically, when $\phi = 0.5\pi$, the energy spectra under the PBC form two closed loops, one of which has $w = +1(-1)$ for Re $(E) < 0$ [Re $(E) > 0$], indicating the corresponding eigenstates ($\Psi$) localized on the left (right) side of the lattice [see Fig. 2(c3) also]. This is the so-called bipolar NHSE [35,45], where eigenstates can localize on both sides of the lattice depending on the eigenenergies. However, the energy spectra under the OBC and PBC overlap for $\phi = 0, \pi$ with $w = 0$, and thus the NHSE disappears [see Figs. 2(a3) and 2(d3) also]. We plot eigenstates at Re$(E) = -0.2$ for different $\phi$ in Fig. 1(b). It is found that eigenstates localize at the right boundary of lattices with $\phi = 0.25\pi$ and $\phi = 0.5\pi$, whereas they show extended distributions for cases of $\phi = 0$ and $\pi$. Thus, the NHSE can be controlled by the effective magnetic flux in the non-Hermitian two-leg Hall ladder.

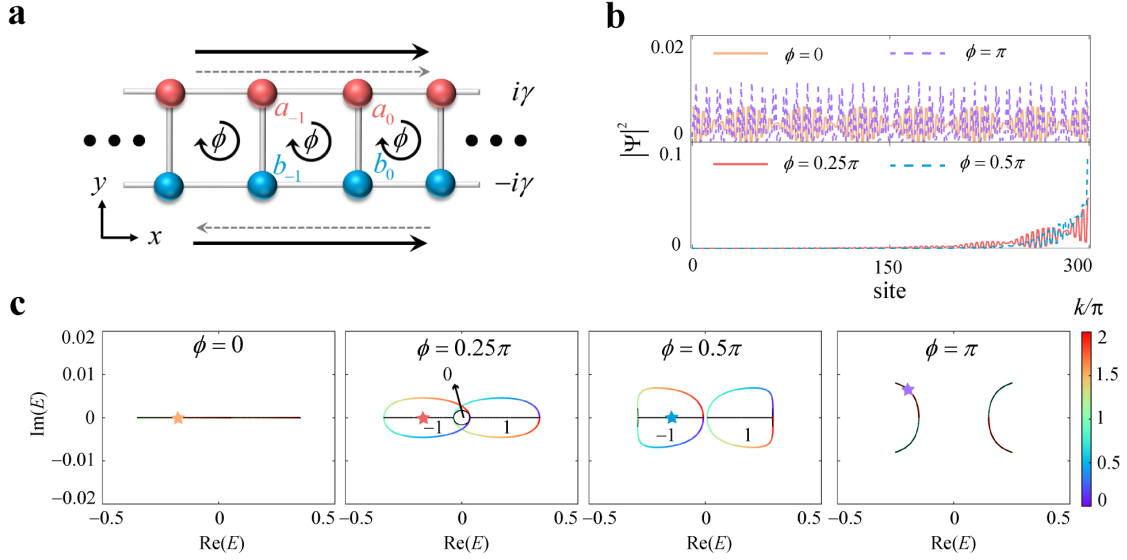

**Fig. 1. Non-Hermitian two-leg Hall ladder with the ACECs and NHSE.** (a) The non-Hermitian two-leg Hall ladder with on-site gain ($i\gamma$) and loss ($-i\gamma$) on legs $a$ and $b$. The inter-leg and intra-leg coupling strengths are $\kappa$ and $t$, respectively. The solid (dashed) arrows denote the directions of the ACECs (CECs) for $\gamma \neq 0$ ($\gamma = 0$). (b) Eigenstates corresponding to one of eigenenergies with $\mathrm{Re}(E) = -0.2$ for different lattices under the OBC with $\phi = 0$ (orange solid line), $\phi = \pi$ (purple dashed line), $\phi = 0.25\pi$ (red solid line), and $\phi = 0.5\pi$ (blue dashed line). (c) Energy spectra under the PBC (colored line) and OBC (black dots) for lattices with $\phi = 0$, $\phi = 0.25\pi$, $\phi = 0.5\pi$ and $\phi = \pi$, wherein the starts denote one of the eigenenergies with $\mathrm{Re}(E) = -0.2$. The numbers in the figures indicate the value of $w$ in different loops. Other parameters are $\kappa = 0.15, t = 0.1$, and $\gamma = 0.01$.

In such a proposed model, the NHSE can be understood from the ACECs, which originate from the interplay between the effective magnetic flux and on-site gain and loss. In Figs. 2(a1)-2(d1), we plot band structures for different choices of $\phi$ with the ratio of the distribution of eigenstates on the leg $a$, defined as $P_z = \left(|\psi^a|^2 - |\psi^b|^2\right)/\left(|\psi^a|^2 + |\psi^b|^2\right)$, where $\psi^a$ and $\psi^b$ being components of eigenstates on the legs $a$ and $b$, respectively (see Methods). When $\phi = 0.5\pi$, we can see the eigenstates on two legs $a$ and $b$ have the opposite dispersion for a specific eigenenergy, which are the CECs as signatures from the Hermitian topology. For example, for the eigenenergy with $\mathrm{Re}(E) = -0.2$, the current on the leg $a(b)$ propagates along $+\hat{x}(-\hat{x})$ axis, due to the positive (negative) dispersion. However, the CECs are further influenced by the on-site gain and loss. The current on the leg $a$ with $i\gamma$ does not change its direction, as it experiences gain and get increased. However, the contribution on the current on the leg $b$ gradually decays to zero due to the loss; meanwhile, the increasing current on the leg $a$ leaks into the leg $b$ via the coupling between two legs. As a result, the interplay between two contribution trends changes the current direction on the leg $b$ and makes it propagate along $+\hat{x}$ direction. Therefore, the chirality of the CECs breaks down, and the directions of currents along the two legs become the same, which are the ACECs.

We can quantitively describe the ACECs using the current definition [63]

$$J_{a(b)}(E) = \sum_{i=1,2} \int dk \delta(E_i - E)|\langle\psi_i|a(b)\rangle|^2 \frac{\partial \mathrm{Re}[E_i(k)]}{\partial k}, \tag{1}$$

where $i = 1,2$, $\psi_i = (\psi_i^a, \psi_i^b)$ are the eigenstates of the Bloch Hamiltonian $H_k$ (see Methods), $k \in (0, 2\pi]$ is the Bloch wave number, $\partial \text{Re}[E_i(k)]/\partial k$ represents the group velocity, and $\delta(E_i - E)$ is the Dirac delta function which characterizes the density of states. Considering the lifetime of the states, we can express the density of states as

$$\int dk \delta(E_i - E) = \sum_k \frac{1}{\pi} \frac{\text{Im}[E_i(k)]}{\{E_i(k) - \text{Re}[E_i(k)]\}^2 + \{\text{Im}[E_i(k)]\}^2}, \quad (2)$$

where $\text{Im}[E_i(k)]$ is the imaginary part of the eigenenergy which indicates the lifetime of the eigenstates. From Eqs. (1) and (2), we see the sign of $J_{a(b)}(E)$ is determined by the group velocity $\partial \text{Re}[E_i(k)]/\partial k$, projection of eigenstates on the leg $a(b)$ $|\langle \psi_i | a(b) \rangle|^2$, and the sign of $\text{Im}[E_i(k)]$. If $J_{a(b)}(E) > 0$, the current on the leg $a(b)$ propagates along $+\hat{x}$ direction. In contrast, if $J_{a(b)}(E) < 0$, the current propagates along $-\hat{x}$. From Figs. 2(b2) and 2(c2), one can see the ACECs for $\phi = 0.25\pi$ and $\phi = 0.5\pi$. In contrast, when $\phi = 0, \pi$, there are no ACECs, as shown in Figs. 2(a2) and 2(d2).

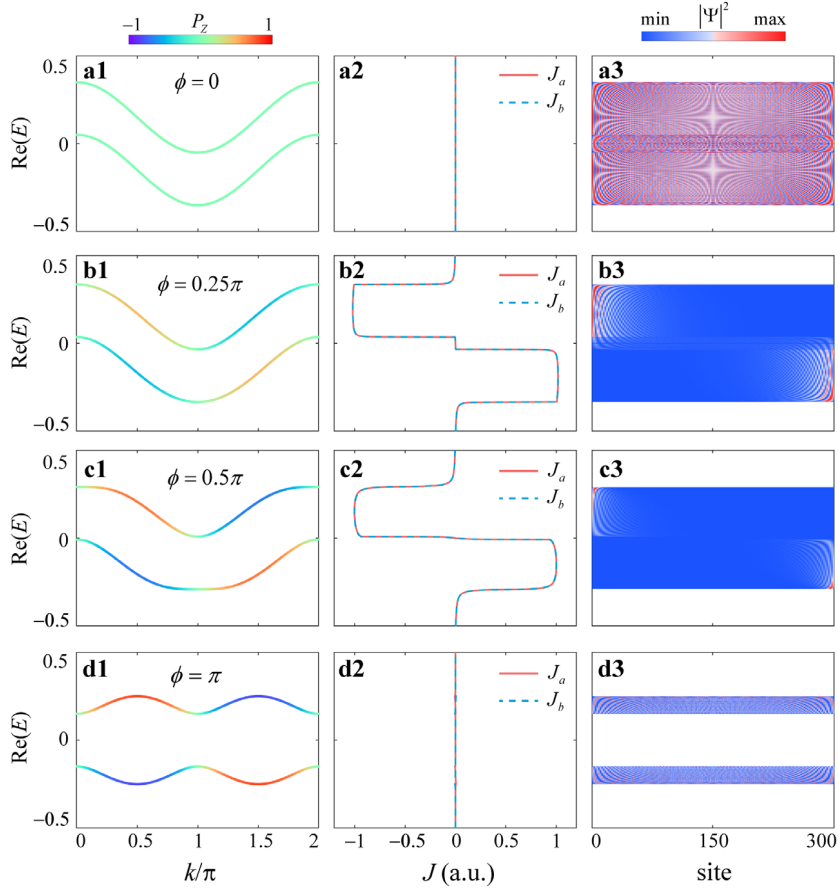

**Fig. 2. Correspondence between the ACECs and the NHSE.** (a1-d1) Band structures for different choices of $\phi$. The color shows the distribution of eigenstates on the leg $a$, defined as $P_z$. (a2-d2) Currents $J_{a,b}$ obtained from Eq. (1). (a3-d3) Distributions of the eigenstates for all eigenenergies under the OBC. (a1-a3) $\phi = 0$, (b1-b3) $\phi = 0.25\pi$, (c1-c3) $\phi = 0.5\pi$, and (d1-d3) $\phi = \pi$. Other parameters are the same as those used for Fig. 1.

To confirm the correspondence between the ACECs and the NHSE, we further plot distributions of the eigenstates for all eigenenergies under the OBC for different $\phi$. The correspondence between the currents $J_{a,b}$ (ACECs) and the eigenstate distributions can be noticed by comparing Figs. 2(a2-d2) and Figs. 2(a3-d3). When $J_{a,b} \neq 0$, the NHSE exists and the direction of the NHSE is consistent with the sign of $J_{a,b}$, i.e., the eigenstates localize at the left (right) boundary of the lattice for $J_{a,b} < 0$ ($J_{a,b} > 0$). The boundary of the localized states agrees well with the current direction for each eigenenergy. For example, when $\phi = 0.5\pi$, we notice $J_{a,b} < 0$ ($J_{a,b} > 0$) for $\text{Re}(E) > 0$ [$\text{Re}(E) < 0$] in Fig. 2(c2), while the eigenstates with $\text{Re}(E) > 0$ [$\text{Re}(E) < 0$] localize at the left (right) boundary of the lattice under the OBC [see Fig. 2(c3)]. When $\phi = 0.25\pi$, there is a range of eigenenergies near $\text{Re}(E) = 0$ with $J_{a,b} = 0$ [see Fig. 2(b2)], and thus the corresponding eigenstates are extended over the lattice sites (no NHSE), as shown in Fig. 2(b3). However, when $\phi = 0$ or $\pi$, the ACECs disappear ($J_a = J_b = 0$) for all eigenenergies, and thus there is no NHSE [see Figs. 2(a1)-2(a3) and 2(d1)-2(d3)].

**Measurements of the ACECs.** To experimentally observe the ACECs, we construct a passive non-Hermitian Hall ladder with on-site pseudo-gain and loss in a frequency dimension. The schematic of the experimental setup is illustrated in Fig. 3(a) (see Supplementary Materials for the detailed experimental setup). Two fiber ring resonators $A$ and $B$ at the same length $L_A = L_B = L$ are coupled by a 2× 2 fiber coupler (beam splitter, BS3) with a coupling ratio of 60:40. In the absence of group-velocity dispersion, each ring resonator supports a series of resonant frequencies $\omega_n^a = \omega_n^b = \omega_n = \omega_0 + n\Omega$ [see Fig. 3(b)], where $\omega_0$ is a reference resonant frequency, $\Omega$ is the free spectral range (FSR), and $n = 0, \pm 1, \pm 2, ...$ is the index of resonant frequency modes. Frequency modes in different rings that have the same frequency are coupled by the fiber coupler at the coupling strength $K$. Besides, inside each ring, the nearest-neighbor frequency modes are coupled by phase electro-optic modulators (EOMs), with the phase modulation form $J_A(t) = g\cos(\Omega_R t + \phi_a)$ and $J_B(t) = g\cos(\Omega_R t + \phi_b)$, respectively. Here $g$ is the modulation amplitude, $\phi_a, \phi_b$ are the modulation phases, and $\Omega_R = \Omega$ is the resonant modulation frequency. In this architecture, resonant frequency modes $\omega_n^a$ and $\omega_n^b$ represent the lattice sites $a_n$ and $b_n$, and therefore a synthetic two-leg Hall ladder is constructed, with the effective magnetic flux $\phi = \phi_b - \phi_a$ in each plaquette [Fig. 3(b)]. The dissipation of the legs $a$ and $b$ are $\gamma_a$ and $\gamma_b$, respectively. Here $\gamma_b$ can be adjusted by an additional amplitude EOM (EOM3) in ring $B$, so we can make this two-leg lattice have pseudo-gain and loss lattice by taking $\gamma_a = -\gamma + \bar{\gamma}$ and $\gamma_b = \gamma + \bar{\gamma}$ with $\bar{\gamma} = (\gamma_a + \gamma_b)/2$ being the average loss.

The Hamiltonian of the non-Hermitian synthetic lattice in $k_f$-space (see Supplementary Materials) is

$$H_{k_f} = \begin{bmatrix} -i\gamma_a + g\cos(k_f\Omega + \phi_a) & K \\ K & -i\gamma_b + g\cos(k_f\Omega + \phi_b) \end{bmatrix} = -i\bar{\gamma}I + H_k. \quad (3)$$

Here $k_f$ is the wave vector that is reciprocal to the frequency dimension, and thus acts as the time variable [53], $K$ is the coupling strength between the two legs in the frequency dimension, $I$ denotes the identity matrix. Thus, $H_{k_f}$ is equivalent to the Bloch Hamiltonian $H_k$ (see Methods) except for a global loss $\bar{\gamma}$. Since the non-Hermitian topological properties are not affected by the global loss [21], our experimental configuration can be used to observe the ACECs discussed previously.

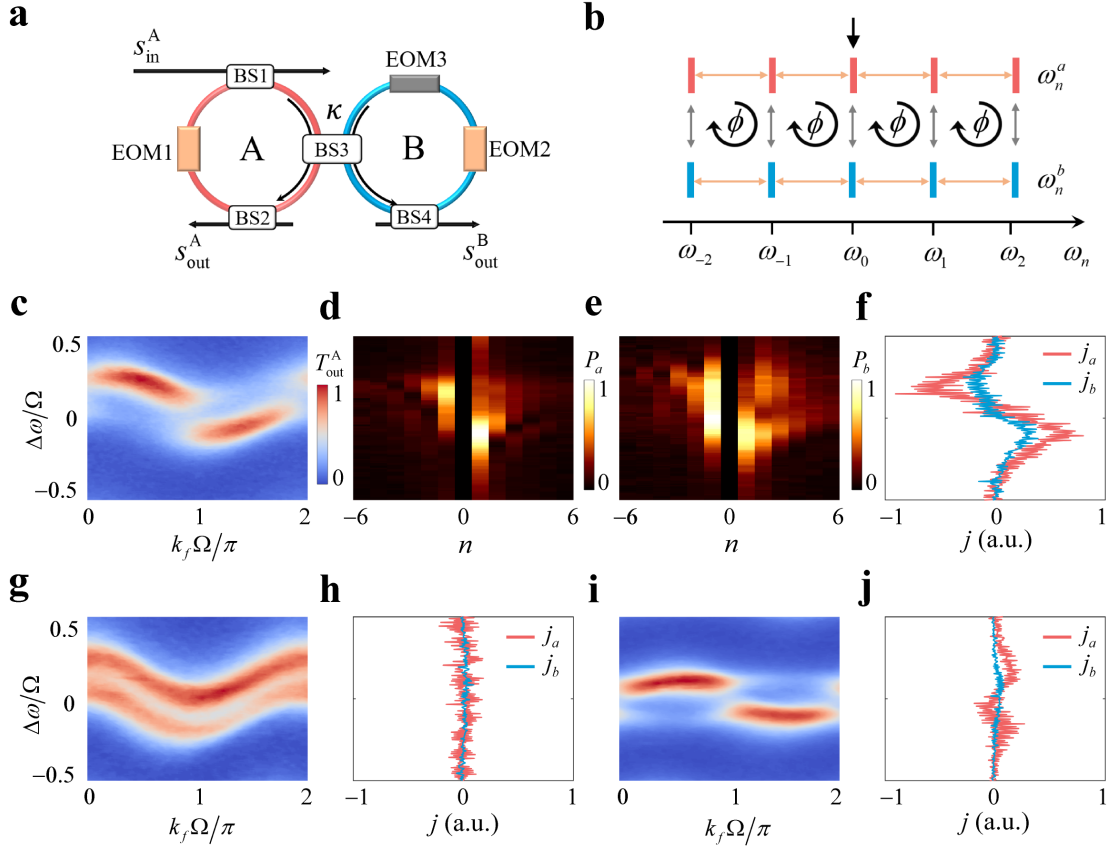

**Fig. 3. Experimental observation of the ACECs** (a) Configuration of two coupled fiber ring resonators under multiple modulations. The input laser field with amplitude $s_{in}^A$ is injected into ring $A$ by the fiber coupler (beam splitter 1, BS1) from the input waveguide. To probe the lattice, two output waveguides are connected to both rings $A$ and $B$ by BS2 and BS4. Two EOM1 and EOM2 with phase modulations $J_A(t)$ and $J_B(t)$ are placed in both resonators to induce the nearest-neighbor couplings in each leg. Another EOM3 with amplitude modulation in ring $B$ is used to adjust the loss so as to tune the dissipation $\gamma_b$ of the legs $b$. (b) The synthetic Hall ladder lattice in the frequency dimension from the experimental configuration in (a). The input laser highlighted by the arrow excites the mode $\omega_0^a$. (c, g, i) Measured projected band structures for $\phi = 0.5\pi$ (c), $\phi = 0$ (g), and $\phi = \pi$ (i). (d, e) Measured steady-state normalized mode distributions $P_a(n)$ (d), $P_b(n)$ (e) for $\phi = 0.5\pi$. (f, h, j) Measured $j_c$ for $\phi = 0.5\pi$ (f), $\phi = 0$ (h), and $\phi = \pi$ (j). Other experimental parameters roughly refer to $K = 0.1\Omega, g = 0.1\Omega, \gamma = 0.04\Omega$, and $\bar{\gamma} = 0.1\Omega$.

We first obtain the band structures from the time-resolved transmission measurements [55]. A tunable continuous-wave laser with amplitude $s_{in}^A$ and a frequency detuning $\Delta\omega$ respect to the reference frequency $\omega_0$ injects into ring $A$ through the input waveguide [see Fig. 3(a)]. The normalized drop-port transmission from ring $A$ can be expressed as (see Supplementary Materials)

$$T_{out}^A(t = k_f; \Delta\omega) = \left|\frac{s_{out}^A}{s_{in}^A}\right|^2 = \sum_{i=1,2} \frac{\gamma_A^2 \left|\psi_{k_f,i}^A\right|^4}{\left|\Delta\omega - \varepsilon_{k_f,i}\right|^2}, \quad (4)$$

where $s_{out}^A$ denotes the drop-port output field amplitude, $\gamma_A$ represent the coupling rate that all modes couple to the input waveguide, $\varepsilon_{k_f,i} = E_i - \bar{\gamma}$ represent the eigenvalues of two-band $H_{k_f}$, and $\psi_{k_f,i}^A = \psi_i^a$ denotes the distribution of corresponding eigenstates on the leg $a$. Equation (4) indicates the projected band structure on the leg $a$ that reads out from ring $A$. The projected band structures on the leg $b$ can also be obtained following the similar procedure by exciting the ring $B$. In Figs. 3(c,g,i), we plot the measured projected band structures on the leg $a$ for different effective magnetic flux $\phi = 0.5\pi, 0, \pi$ with $g = 0.2\Omega, K = 0.1\Omega, \gamma = 0.04\Omega, \bar{\gamma} = 0.1\Omega$ that are taken from experimental parameters, which are consistent with the tight-binding analysis (see Fig. 2). Simulations also show the same band structures and ACECs, which are shown in Supplementary Materials.

To measure the currents in the frequency dimension, we use the input laser to excite the mode $\omega_0^a$ (i.e., the 0-th site on the leg $a$) in the frequency dimension. The currents on the legs $a$ and $b$ are governed by the $J_{a,b}$ in Eq. (1), and are finally balanced by the dissipation in the steady-state limit. To quantify the currents, we, therefore, define the steady-state current as [56]

$$j_{a(b)} = \sum_{n>0} P_{a(b)}(n) - \sum_{n<0} P_{a(b)}(n), \tag{5}$$

where $P_{a(b)}(n)$ represent the steady-state mode distributions of mode $\omega_n^{a(b)}$ on the leg $a(b)$. Therefore $j_{a(b)} \neq 0$ indicate a unidirectional current in the frequency dimension [56], and ACECs can be demonstrated if the signs of $j_a$ and $j_b$ are the same. Furthermore, we can obtain the steady-state mode distributions by using the heterodyne detection method [56]. In experiments, we choose the frequency shift of the input laser by $\delta\omega = 200$ MHz using an acousto-optic modulator and interfere it with the drop-port output field to obtain the interfering field. The mode distribution $P_{a(b)}(n)$ then can be obtained by a fast Fourier transform (FFT) of the interfering field. The frequency of the input laser is scanned through the whole band structure, so we can obtain $P_{a(b)}(n)$ corresponding to energies of all bands. Figures 3(d) and 3(e) show $P_a(n)$ and $P_b(n)$ for the non-zero magnetic flux $\phi = 0.5\pi$. We can see $P_a(n)$ and $P_b(n)$ are both biased to $n < 0$ ($n > 0$) for $\Delta\omega > 0$ ($\Delta\omega < 0$), indicating the ACECs along $-\hat{x}$ ($+\hat{x}$) or lower- (higher-) frequency dimension. The corresponding $j_a$ and $j_b$ are shown in Fig. 3(f), where we observe $j_{a,b} < 0$ ($j_{a,b} > 0$) for $\Delta\omega > 0$ ($\Delta\omega < 0$), which are the hallmark of the ACECs in the synthetic non-Hermitian Hall ladder lattice. Note the amplitude of $j_a$ is larger than that of $j_b$, which is due to the initial excitation on the leg $a$. The signs of $j_a$ and $j_b$, however, are the same, which is the evidence of the ACECs. The existence of ACECs gives the key signature of the NHSE (see the previous explanations in Fig. 2), resulting from the interplay between the effective magnetic flux and the onsite gain/loss. The numerical simulations of the steady-state distributions and currents from the Floquet analysis agree well with the experimental observations (see Supplementary Materials). Besides, we show the projected band structures and currents for $\phi = 0, \pi$ in Figs. 3(g)-3(j) and observe $j_a = j_b = 0$, which shows no ACECs. Therefore, there is no NHSE in these cases.

**Measurements of topological windings of the band structures.** Besides the ACECs, the topological winding of the energy bands can provide another evidence of the NHSE. From the projected band structures of the lattice, we can obtain the real and imaginary parts of eigenvalues, $\text{Re}(E)$ and $\text{Im}(E)$ (see Methods). We again take the case of $\phi = 0.5\pi$, and showcase the experimentally obtained $T_{\text{out}}^A$ at $k_f \Omega = \pi$ as a function of $\Delta\omega$ in Fig. 4(a4), from which the $\text{Re}(E)$ and $\text{Im}(E)$ can be extracted from fitting the locations and the linewidths of the two peaks. Using this method, we plot the extracted $\text{Re}(E)$ and $\text{Im}(E)$ for the entire first Brillouin zone with $k_f \in (0, 2\pi/\Omega]$ in Fig. 4(a2) and 4(a3). By plotting the $\text{Re}(E)$ and $\text{Im}(E)$ in the complex energy plane, we obtain the band windings, which exhibit two closed loops with the corresponding winding number $\pm 1$. Therefore, the nontrivial windings demonstrate the existence of the NHSE in this non-Hermitian two-leg ladder model under the effective magnetic flux $\phi = \pi/2$. The existence of the NHSE is the result of the interplay between the magnetic flux and the on-site gain/loss. For example, we consider the Hermitian case with $\gamma = 0$ and all other parameters being the same. The topological windings of band structures, $\text{Re}(E)$ and $\text{Im}(E)$, as well as $T_{\text{out}}^A$ are shown in Fig. 4(b1)-4(b4). We see that the $\text{Re}(E)$ and $\text{Im}(E)$ in the complex energy plane form a line, indicating the absence of the NHSE.

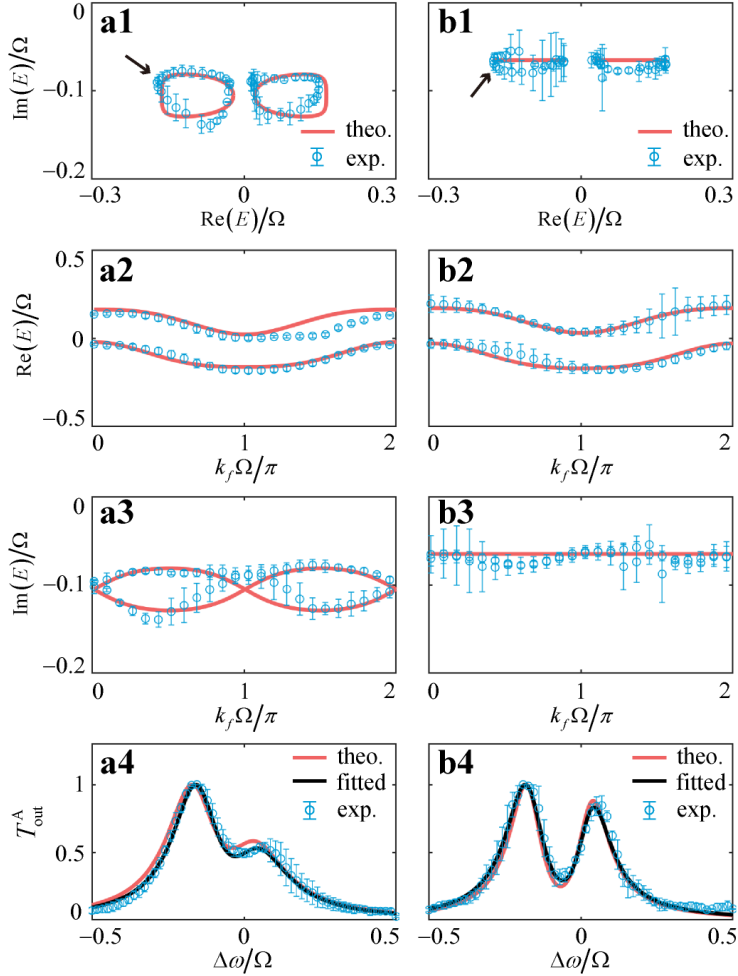

**Fig. 4. Experimental measurements of the topological windings of energy bands in the complex energy plane.** (a1, b1) Experimental and theoretical topological windings in the complex energy plane with $\phi = 0.5\pi$ for non-Hermitian (a1) and Hermitian (b1) cases. Arrows indicate the corresponding slices with $k_f\Omega = \pi$, which are shown in (a4) and (b4). Experimental measurements: circles with error bars; theoretical calculations: lines. (a2, b2) Experimental and theoretical $\text{Re}(E)$ for non-Hermitian (a2) and Hermitian (b2) cases. (a3, b3) Experimental and theoretical $\text{Im}(E)$ for non-Hermitian (a3) and Hermitian (b3) cases. (a4, b4) Experimental and theoretical $T_{\text{out}}^A$ as a function of $\Delta\omega$ at $k_f\Omega = \pi$ for non-Hermitian (a4) and Hermitian (b4) cases. The curve of Eq. (10) (red lines) is used for fitting (see Methods). The parameters in the non-Hermitian case (a1-a4) are the same as those in Fig. 3(c). The parameters in the Hermitian case (b1-b4) are the same as those in (a1-a4) except for $\gamma = 0$.

**Observation of the ACECs in the lattice under the long-range coupling.** Our proposed model can be further generalized to the synthetic lattice with the long-range coupling and study the NHSE therein. Here, we introduce the next-near-neighbor (NNN) coupling on the leg $a$ in the synthetic two-leg lattice by replacing the phase modulation in the ring $A$ by $J_A(t) = g\cos(\Omega t + \phi_a) + g'\cos(2\Omega t + \phi'_a)$, with $g'$ and $\phi'_a$ being the long-range modulation amplitude and phase. This configuration builds a synthetic lattice model with the three-dimensional (3D) structure, having an additional locally non-uniform effective magnetic flux $\pm\Theta$, where $\Theta = 2\phi_a - \phi'_a$ [see Fig. 5(a)]. Such a 3D synthetic lattice actually holds effective magnetic monopoles, as there exists locally non-zero magnetic flux through enclosing surfaces here. The theoretical band structure, energy spectra, currents, and distributions of the eigenstates for all eigenenergies for $\Theta = 0.5\pi$ and $\phi = 0$ are shown in Fig. 5(b-e). From Fig. 5(c), we see the energy spectra under the PBC form several loops, which are different from those under the OBC, thus indicating the existence of the NHSE. The directions of the windings of energy spectra under the PBC are consistent with those of the NHSE [see Fig. 5(e)]. Besides, from Fig. 5(d), we can see the signs of $J_a$ and $J_b$ are the same for a specific eigenenergy, which indicates the ACECs with a total current $J = J_a + J_b$. The directions of the ACECs (the sign of $J$) under the PBC are also consistent with the existence of the NHSE under the OBC [see Fig. 5(e)], where the eigenstates with $|\text{Re}(E)| \in (0,0.1)$ localize at the right boundary of the lattice, and the eigenstates with $|\text{Re}(E)| \in (0.1,0.2)$ localize at the left boundary of the lattice.

In the experiment, we measure the band structure and the currents, which are shown in Fig. 5(f) and 5(g), demonstrating the ACECs by the signature of $j_{a,b} > 0$ at $\Delta\omega \approx 0$ and $j_{a,b} < 0$ at $\Delta\omega/\Omega \approx \pm 0.1$. It is worth noting that this model can achieve a higher spectral winding number ($|w| \neq 0,1$) and probe the corresponding NHSE in this case. We can see $w = -2$ in the vicinity of $\text{Re}(E) = 0$ [Fig. 5(c)] and the corresponding current $J$ is nearly twice as the amplitudes of those eigenenergies with $w = \pm 1$ [Fig. 5(d)]. More interestingly, we notice that multiple local effective magnetic fluxes can be introduced in our model by tuning $\phi_a, \phi_b, \phi'_a$ to achieve $\phi, \Theta \neq 0$, where the ACECs and the corresponding NHSE under the different fluxes in this 3D lattice can be studied (see Supplementary Materials). In addition, we can add the NNN couplings on both legs, which can give another degree of freedom of the effective gauge flux around the closed loop among sites $b_0 \rightarrow b_1 \rightarrow b_{-1} \rightarrow b_0$ as well as closed loops among sites $a_{-1} \rightarrow a_1 \rightarrow b_1 \rightarrow b_{-1} \rightarrow a_{-1}$ and $a_0 \rightarrow a_2 \rightarrow b_2 \rightarrow b_0 \rightarrow a_0$. It is found that this model can exhibit more complex ACECs and NHSE, which are discussed in Supplementary Materials.

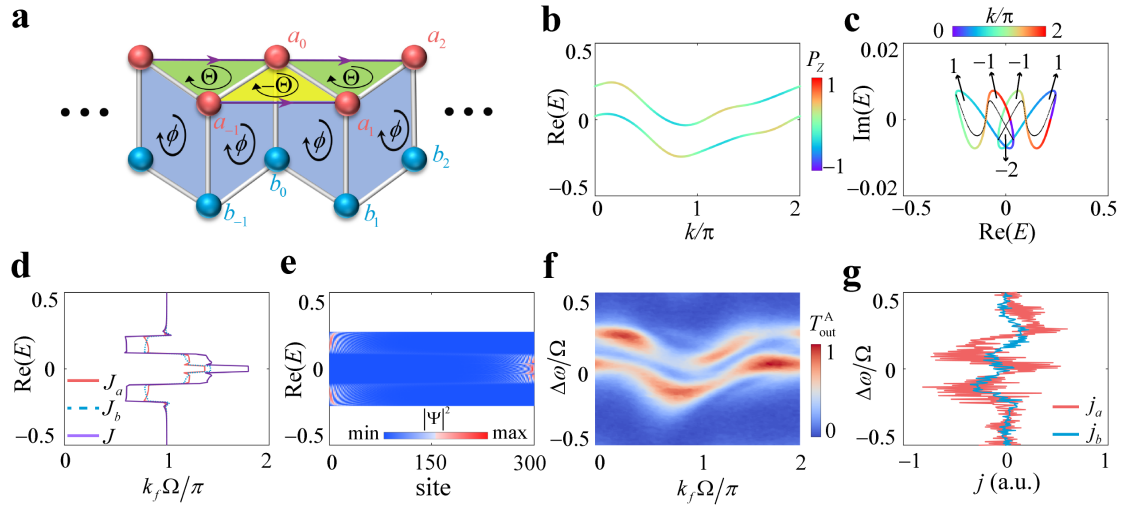

**Fig. 5. Experimental observation of the ACECs in the synthetic lattice with additional long-range couplings.** (a) The synthetic non-Hermitian Hall ladder lattice including NNN coupling $t'e^{\pm i\phi'_a}$ on leg $a$. Such a configuration supports a 3D synthetic lattice and also introduces an additional local non-uniform effective magnetic flux $-\Theta(\Theta)$ in the up (down) triangle. (b) The corresponding band structure for the lattice with parameters $\Theta = 0.5\pi, \phi = 0, t = 0.06, \kappa = 0.1, t' = 0.04$, and $\gamma = 0.02$. (c) Energy spectra under the PBC (colored line) and OBC (black dots). The numbers in the figures indicate the value of $w$ in different loops. (d) The calculated currents $J_{a,b}$ and the total currents $J$. (e) Distributions of the eigenstates for all eigenenergies. (f) Experimentally measured projected band structure. (g) The measured currents $j_{a,b}$. The experiments give parameters $K = 0.1\Omega, g = 0.12\Omega, g' = 0.04\Omega, \bar{\gamma} = 0.07\Omega$, and $\gamma = 0.02\Omega$.

## Discussion

To conclude, we have theoretically proposed and experimentally observed non-Hermitian ACECs in the synthetic frequency dimension in the optical regime, based on a passive two-leg Hall ladder that holds the interplay between the on-site pseudo gain and loss and the effective magnetic flux. The currents on the two legs having the same direction, which are the hallmark of ACECs, are experimentally demonstrated by the steady-state frequency mode distributions under the initial excitation on one frequency site. Such ACECs, derived from the bulk band structures under the PBC, predict the corresponding NHSE under the OBC, which is also confirmed by measuring the topological windings of the energy bands. Our results thus provide the signatures for probing the NHSE without applying an open boundary. Furthermore, we consider the influence of the long-range couplings in this lattice model and prove the existence of ACECs in a synthetic 3D structure that supports higher spectral winding numbers. As a side note, the observed non-Hermitian ACECs here are also related to the persistent currents proposed in Ref. [49], which have been considered to be one of the physical origins of the NHSE. The non-zero ACECs related to each eigenenergy in our two-leg model thus demonstrate the existence of nonvanishing currents, and moreover, can predict the existence of the NHSE for a specific eigenenergy, which is absent in previous study [46,47,49]. Our work demonstrates a new mechanism leading to ACECs in experiments, which can be implemented at other electromagnetic wavelength regime and may trigger future interest in combining effects from topology and non-Hermiticity. The capability for constructing complex 3D synthetic lattices may also be further extended to study physics in higher dimensions, such as antichiral surface states [64,65] in the optical regime. The phenomena of ACECs here might find potential applications in robust photonic devices desiring unidirectional frequency conversion and selective lasing at higher/lower frequency modes.

## Methods

### Hamiltonians and spectral winding number of the model

The Hamiltonian that describes the lattice model in Fig. 1(a) is

$$H = \sum_n i\gamma(a_n^\dagger a_n - b_n^\dagger b_n) + \kappa \sum_n (a_n^\dagger b_n + b_n^\dagger a_n) + [(te^{-i\phi/2} a_n^\dagger a_{n+1} + te^{i\phi/2} b_n^\dagger b_{n+1}) + h.c.], \quad (6)$$

where $a_n^\dagger(b_n^\dagger)$ and $a_n(b_n)$ are creation and annihilation operators of $n$-th lattice site in legs $a$ and $b$, respectively. $\kappa$ is the coupling strength between sites on two legs and $t$ describes the nearest-neighbor hopping strength between two sites on each leg. The difference between hopping phases on two legs gives the effective magnetic flux $\phi$ for photons. If the lattice is infinite, the corresponding Bloch Hamiltonian in the momentum space ($k \in (0, 2\pi]$ in the first Brillouin zone) is

$$H_k = \begin{bmatrix} i\gamma + 2t\cos\left(k - \frac{\phi}{2}\right) & \kappa \\ \kappa & -i\gamma + 2t\cos\left(k + \frac{\phi}{2}\right) \end{bmatrix}, \quad (7)$$

which gives the band structure

$$E_{1,2}(k) = 2t\cos k \cos\frac{\phi}{2} \pm \sqrt{\left(2t\sin k \sin\frac{\phi}{2} + i\gamma\right)^2 + \kappa^2}, \quad (8)$$

and corresponding eigenstates $\psi_{1,2} = (\psi_{1,2}^a, \psi_{1,2}^b)$ with $\psi_{1,2}^a$ and $\psi_{1,2}^b$ being components on the legs $a$ and $b$, respectively. The NHSE origins from the point-gap topology of band structures, which can be characterized by the spectral winding number $w$ [50]

$$w = \sum_{i=1,2} \int_0^{2\pi} \frac{dk}{2\pi} \partial_k \arg[E_i(k) - \epsilon], \quad (9)$$

where $\epsilon$ is a reference energy in the complex energy plane. The nontrivial $w \neq 0$ indicates the existence of the NHSE, and the sign of $w$ determines the direction of the NHSE.

**Experimental details**

Two optical fiber ring resonators ($A$ and $B$) with the same length $L = 11.6$ m and the corresponding free-spectral range (FSR) $\Omega = 2\pi \times 17.6$ MHz are coupled through a $2 \times 2$ polarization-maintaining fiber coupler with the coupling ration 60:40 (see the detailed experimental setup in Supplementary Materials). A low-noise telecommunications C-band continuous-wave (CW) laser with a 200 kHz linewidth centered at 1550.92 nm is injected into ring $A$ through a $2 \times 2$ fiber coupler with a coupling ratio 99:1. The other two 99:1 fiber coupler couples couple 1% of the laser insider two rings to the drop-port waveguide for heterodyne beating with the laser source and band structure measurements. The laser frequency can be scanned over 30 GHz by applying a ramp signal with amplitude V = 1.0 Vpp and frequency 100 Hz on the frequency module. Two identical lithium niobate phase electro-optic modulators (EOMs) with 10 GHz bandwidth are driven by two arbitrary waveform generators (bandwidth 200 MHz). The phase difference between the two modulators can be adjusted through software in the computer. Two semiconductor optical amplifiers (SOAs) are used to compensate for the losses in both rings to obtain high-quality factors. Two dense-wavelength division multiplexing (DWDM) band-pass filters (Channel 33, center wavelength 1550.92 nm) are utilized to effectively suppress the amplified spontaneous emission noise emanating from the SOA. Two polarization controllers in both rings ensure that the polarization orientation of the laser in rings matches the principle axis of EOMs. An additional amplitude EOM which is controlled by a DC bias in the linear regime of the lithium-niobate-waveguide Mach-Zehnder interferometer (MZI) fringe is incorporated into ring $B$ to add an additional loss of the ring $B$. Both fiber rings are coupled to through- and drop-ports to enable an independent calibration of the FSR of each ring when the coupling between the two rings is absent. The two drop-port signals are directly sent to two fast InGaAs photodiodes (850 to 1650 nm with 10 GHz bandwidth) after optical amplification with an erbium-doped optical fiber amplifier (with a maximum gain of 12 dB) and then are sent to the oscilloscope (5 G samples/s with 1 GHz bandwidth) for band structure measurement. Meanwhile, two parts of the laser source are frequency-shifted by 200 MHz through two acousto-optic modulators (AOMs) and then interfere with the two drop-port signals to enable heterodyne detection of the steady-state frequency mode distributions on the two legs.

**Method for measuring projected band structures**

To obtain the projected band structures, we scan the frequency of the input laser around several FSRs and obtain the drop-port transmission spectra [55]. Then we break the transmission spectra into different time slices with the time window $T = 2\pi/\Omega$, which is the one roundtrip time of ring $A$ or $B$. By stacking up these time slices as a function of the frequency of the input laser, we obtain the projected band structures in Figs. 3(c,g,i) and 5(f).

**Method for obtaining the real and imaginary part of eigenvalues**

To obtain the real and imaginary parts of the eigenvalues from the projected band structures, we need to fit the experimentally measured $T_{\text{out}}^A$ at as a function of $\Delta\omega$ to obtain the locations and bandwidths of the two peaks. From Eq. (4), we see the function of $T_{\text{out}}^A$ with $\Delta\omega$ are two Lorentz functions. Then the fitting function can be

$$T_{\text{out}}^A(\Delta\omega) = \left|\frac{R_1}{[\text{Re}(E_1) + i\text{Im}(E_1) - \Delta\omega]} + \frac{R_2}{[\text{Re}(E_2) + i\text{Im}(E_2) - \Delta\omega]}\right|^2, \quad (10)$$

where $R_{1,2}$ are fitting coefficients, $\text{Re}(E_{1,2})$ and $\text{Im}(E_{1,2})$ are the obtained real and imaginary parts of eigenvalues.